\documentstyle[aps,prb,epsfig,twocolumn]{revtex}

\begin{document}

\draft

\twocolumn[\hsize\textwidth\columnwidth\hsize\csname @twocolumnfalse\endcsname


\title{Charge ordering in half-doped Pr(Nd)$_{0.5}$Ca$_{0.5}$MnO$_3$
under magnetic field}
\author{Qingshan Yuan$^{1,2}$ and Thilo Kopp$^1$}
\address{$^1$ Exp VI, Center for Electronic Correlations and 
Magnetism, Universit\"at Augsburg, 86135 Augsburg, Germany\\
$^2$Pohl Institute of Solid State Physics, Tongji University, 
Shanghai 200092, P.R.China}

\maketitle

\begin{abstract}
Recent experiments in Pr(Nd)$_{0.5}$Ca$_{0.5}$MnO$_3$ thin films 
exhibited (multiple) reentrant charge ordering (CO) transitions with change of
{\it temperature} $T$ under fixed magnetic field $H$, 
which are in contrast to the results for the corresponding bulk materials. 
To explain the experimental findings,
a model including the double-exchange mechanism, intersite
Coulomb interaction and electron-phonon coupling is proposed, for which 
the reentrant CO is naturally obtained due to
the temperature dependence of the band-type coherent polaron hopping.
Various results for the CO in the ($H,\ T$) plane
are extensively discussed. The theory is considered to be valid
for both thin films and bulk materials.
\end{abstract}


]

\section{Introduction}
The colossal magnetoresistance manganites R$_{1-x}$A$_x$MnO$_3$ 
(R: rare-earth element like La, Pr, Nd; A: divalent alkali like 
Ca, Sr) and R$_{n-nx}$A$_{1+nx}$Mn$_n$O$_{3n+1}$ ($n=1,2$)
have attracted great interest for both bulk materials and thin films
in recent years.\cite{Salamon}
Due to the complex interplay of charge, orbital, spin, and lattice degrees of 
freedom,\cite{Maezono,Millis,Horsch} intriguing phenomena 
have been discovered. Among them, charge ordering (CO) with alternating 
Mn$^{3+}$ and Mn$^{4+}$ ions in real space is one of the most 
remarkable findings.\cite{Rao}
It has been observed for optimal doping $x=0.5$ in various (R, A) 
compositions, where the concentrations of Mn$^{3+}$ and Mn$^{4+}$
ions are equal, and was also found in the doping region $0.3<x<0.75$, 
depending on R and A.\cite{Imada} The CO may be melted by external 
magnetic fields, which are favorable to a ferromagnetic (FM) metallic state.

\begin{figure}[ht]
\centerline{\epsfig{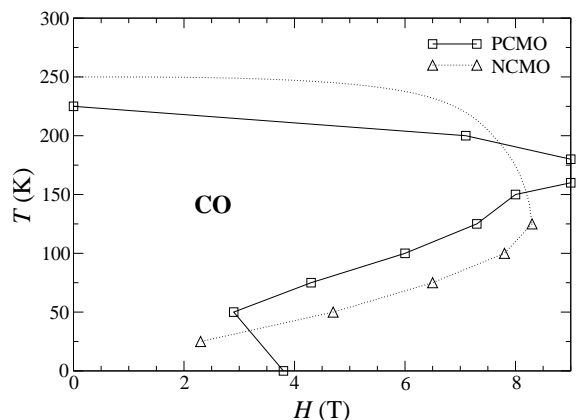}}
\smallskip
\caption{The experimental CO phase diagrams in the ($H,\ T$) plane
for Pr$_{0.5}$Ca$_{0.5}$MnO$_3$ (PCMO) and Nd$_{0.5}$Ca$_{0.5}$MnO$_3$ 
(NCMO) thin films, extracted from Refs. [8] and [11],
respectively. Note that the critical field $H_c$ has been taken as the 
average of the values in the field-increasing and decreasing scans.}
\label{Fig:PDexp}
\end{figure}

The CO in manganite thin films has been studied very recently.
\cite{PreGos,PreSim,Ogimoto,Biswas,Buzin}
Due to strain effect, different properties from those in bulk materials
are expected. Prellier {\it et al.} studied the CO instability under magnetic 
field $H$ in Pr$_{0.5}$Ca$_{0.5}$MnO$_3$ thin films\cite{PreGos,PreSim}
and found a spectacular reduction of the melting field $H_c$ for CO
under tensile strain. This is ascribed to an enhanced electron itineracy.
\cite{PreGos}
On the other hand, the critical $H_c$, as a function of 
temperature $T$, is found to be non-monotonous. 
From the phase diagram in the plane ($H$, $T$),\cite{PreSim} 
which is replotted in Fig.~\ref{Fig:PDexp} by the solid line, 
a more interesting phenomenon, i.e., reentrant CO transitions with 
change of $T$ under fixed $H$, can be clearly seen.
For example, at $H=6$ T, the CO appears at about $T=210$ K 
but later melts at $T=100$ K. The profile for $H=3.5$ T is even more
remarkable: the CO appears at about $T=220$ K, and later melts at 
$T=60$ K, but finally reappears at $T=20$ K, i.e., 
multiple (twice) reentrant CO transitions are expected through the whole
temperature region. A similar phase diagram with possible single reentrant 
behavior was also reported 
for Nd$_{0.5}$Ca$_{0.5}$MnO$_3$ thin films, \cite{Buzin} 
and is reproduced in Fig.~\ref{Fig:PDexp} by the dotted line.
In contrast, no obvious reentrant CO was found for the two
corresponding bulk materials.\cite{Tokura,Tomioka,Tokunaga,Respaud,Supp} 
Naturally, a theoretical investigation of the 
experimental findings is required.

It should be pointed out that the reentrant CO under magnetic field
was clearly observed in bulk R$_{1-x}$Ca$_{x}$MnO$_3$ (R=Pr, Nd) for 
doping {\it away from}
$0.5$, e.g., $0.3\le x< 0.45$ for R=Pr.\cite{Jirak,Asami,Tomioka,Tokunaga} 
Although this issue is not fully understood either, it was argued 
by Khomskii {\it et al.} that phase separation into an electron-rich 
region (due to extra Mn$^{3+}$ ions) and a CO region, as that 
for $x=0.5$, is a relevant mechanism.\cite{Khomskii,Kagan} 
On the other hand, even in the half-doped ($x=1/2$) case, 
(multiple) reentrant CO was found before in
bi-layered manganite LaSr$_2$Mn$_2$O$_7$ 
under zero magnetic field.\cite{Kimura,Chatterji,Dho,Li}
A mechanism for this scenario was proposed by us.\cite{Yuan} 
The idea is to ascribe the CO to the competition of electronic 
kinetic energy and intersite Coulomb interaction, 
and simultaneously take electron-phonon (e-p)
coupling into account, i.e., consider polaron ordering. 
Due to the inherent temperature-dependent band-type polaron motion,
the kinetic energy may be small compared to the Coulomb energy
at some temperatures (favoring CO), but dominates at 
other temperatures (favoring the homogenous state) 
so that a reentrant behavior may appear.
Here, the same idea will be generalized to study the CO 
under magnetic field. The experimental phase diagrams are interpreted 
and, more generally, various results for
CO under magnetic field for half-doped manganites are discussed. 

Before we start to work out the details it deserves to further clarify the 
validity of the above proposed theory.
Actually, we notice that there are several alternative models to study
CO (and concomitant orbital and magnetic orderings) in half-doped
manganites currently.
A natural model is to take the intersite Coulomb interaction 
as the driving force for CO. This modelling was adopted by many authors 
\cite{Yuan,Koshibae,Mishra,Lee,Yu,Mutou,Jackeli,Khomskii,Kagan}
and also used for a discussion of other materials, e.g., the Verwey 
transition in magnetite Fe$_3$O$_4$.\cite{Cullen} 
Another kind of model\cite{Brink,Solovyev} emphasizes the role of 
CE-type antiferromagnetic (AFM) spin order concomitant with CO, 
which was found in e.g., Nd$_{0.5}$Sr$_{0.5}$MnO$_3$.\cite{Tokura} 
As derived by van den Brink {\it et al.},\cite{Brink}
the on-site Coulomb interaction alone between electrons in
different orbitals will lead to CO. And the experimental 
($\pi$,$\pi$,0)-type CO, i.e., the checkerboard CO forms in the $xy$-plane 
and stacks in $z$-direction, can be explained within the same
picture. However, this model is not well suited 
for Pr(Nd)$_{0.5}$Ca$_{0.5}$MnO$_3$, where
the spins are not yet ordered when CO appears.
\cite{Tokura,Tomioka,Tokunaga,Respaud}
Recently, a purely Jahn-Teller (JT) phononic model, even without direct
electron correlations, was proposed by Yunoki {\it et al.}\cite{Yunoki,Hotta} 
By treating the e-p coupling adiabatically,
they have constructed the phase diagram of charge, orbital 
and magnetic orderings with change of JT coupling strength and 
AFM superexchange of the core spins. 
This model seems not to be generally accepted at the moment.\cite{Kagan}
Without relying on peculiar features of concrete materials 
(e.g., magnetic structure), the intersite Coulomb interaction,
which is always present in reality, is regarded as the origin of CO as done
by most authors. And on the other hand, 
another aspect of e-p coupling, i.e., polaron formation with
dynamic phonons is addressed. Experimentally it has been verified
that the CO should be viewed as 
the formation of a polaron lattice.\cite{LiUe,Zhao,Alex} 
A strong evidence is the isotope effect,
where the CO transition temperature $T_{co}$ was changed by 
replacing $^{16}$O with $^{18}$O.\cite{Zhao}
Therefore, as the simplest model to study CO in manganites, the e-p coupling
is indispensable, besides the intersite Coulomb interaction.
This improves the one proposed by Khomskii where only the latter is 
included.\cite{Khomskii}
Indeed, the model with both e-p and intersite Coulomb interactions
has been successfully used to
explain the experimental isotope effect,\cite{Yu}
renormalization of sound velocity around $T_{co}$,\cite{Lee}
and reentrant CO in LaSr$_2$Mn$_2$O$_7$ mentioned above.\cite{Yuan}
Also, it is believed to have grasped much of the physics of the
CO transitions in Pr(Nd)$_{0.5}$Ca$_{0.5}$MnO$_3$, addressed in this paper.

\section{Model Hamiltonian and technical treatments}

On the basis of the considerations in the introductory section,
we write the Hamiltonian as
\begin{eqnarray}
\label{H}
H & = & -t\langle \cos (\theta /2) \rangle \sum_{\langle ij \rangle}
(c_i^{\dagger}c_j + {\rm h.c.}) + V\sum_{\langle ij \rangle} n_i n_j
\nonumber\\
& & +g\sum_i n_i(b_i+b_i^{\dagger}) + 
\omega \sum_i b_i^{\dagger}b_i \ .
\end{eqnarray}
The first term is the double-exchange (DE) model, where $t$ is the 
nearest-neighbor (n.n.) hopping and $\theta$ is the angle between two n.n. 
core spins which is assumed uniform or an average on $\theta_{ij}$. 
$V$ is the intersite Coulomb
interaction and $n_i=c_i^{\dagger}c_i$ is the electron number operator.
Irrespective of the details for various types of e-p coupling 
(e.g., JT, breathing, etc.), we have used a simple on-site Holstein-type 
model with dispersionless phonons\cite{Roder,Green} as shown by the third term 
to describe an essential effect of e-p coupling.
Then $b_i,\ b_i^{\dagger}$ are local 
phonon operators at site $i$, $g$ is the e-p coupling constant and $\omega$ 
is the Einstein phonon frequency. The band is half-filled, i.e., 
one electron per two sites. 
Note that the effect of external magnetic field
$H$ has been implied in the factor $\langle \cos (\theta /2) \rangle$,
which becomes larger with increasing $H$ due to enhanced FM order 
(i.e., $\theta \rightarrow 0$).

In the following, we consider the Hamiltonian (\ref{H}) in the 
two-dimensional (2D) case since
the checkerboard CO was found only in the basal plane as mentioned above.
\cite{Supp2D}
Moreover, we mainly focus on the qualitative results in this work, thus no
sophisticated techniques will be elaborated. The quantitative aspects of the 
results obtained should not be trusted literally, 
but they are qualitatively correct. 
Throughout the paper we set $\hbar=k_B=1$ and $t$ is taken as the
energy unit.

First, the usual Lang-Firsov (LF) transformation is acted upon
Hamiltonian (\ref{H}), i.e., $\tilde{H}=UHU^{\dagger}$ with
$$
U=\exp [-(g/\omega)\sum_i n_i (b_i-b_i^{\dagger})] \ .
$$
Then, thermally averaging on the phonon states, one obtains the following
effective electronic Hamiltonian:
\begin{eqnarray}
H_{\rm eff} & = & -\tilde{t} \sum_{\langle ij \rangle}
(c_i^{\dagger}c_j + {\rm h.c.}) + V\sum_{\langle ij \rangle} n_i n_j 
\label{Heff}
\end{eqnarray}
with the renormalized hopping $\tilde{t}=
t\langle \cos (\theta /2) \rangle P(T)$.
Here 
$$P(T)=\exp [-({g\over \omega})^2\coth ({\omega \over 2T})]$$ 
is the polaron narrowing factor and a constant 
$-(g^2/\omega)\sum_i n_i^2$ is ignored in 
Hamiltonian (\ref{Heff}). 
Thus the effect of e-p coupling is contained exclusively in $P(T)$.

\begin{figure}
\centerline{\epsfig{file=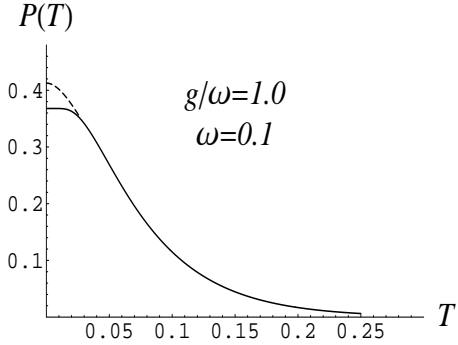,width=6cm,height=4.5cm}}
\medskip
\caption{$P(T)$ as a function of $T$ for $g=\omega=0.1$ (solid line). The 
dashed line schematically shows the result 
with acoustic phonons (see text).}
\label{Fig:PT}
\end{figure}

Note that $P(T)$ has a strong temperature dependence, which is plotted in 
Fig.~\ref{Fig:PT}. Although the narrowing effect is overestimated by the 
LF method for $\omega/t \ll 1$, i.e., the absolute value of $P(T)$ 
is too small, the $T$-dependent property is qualitatively correct, i.e.,
$P(T)$ increases monotonically with decreasing $T$ because of
smaller and smaller phonon occupation number. Due to the same
reason, $P(T)$ is seen to be nearly flat at very low temperatures, about
$T/\omega <0.2$, where the phonons are hardly excited. 
If acoustic phonons are considered, as 
in some references,\cite{AlexBrat,LeeMin,Lee,Yu} 
a rapid variation of $P(T)$ with $T$ should be
still possible up to $T=0$, as schematically shown by the dashed line
in Fig.~\ref{Fig:PT}.
We emphasize here that the $T$-dependent $P(T)$ is the essence of 
our theory, which may render the CO to melt 
with {\it decreasing} $T$ by quickly enhancing 
the hopping $\tilde{t}$. In principle, the thermal average 
$\langle \cos (\theta /2) \rangle$ is also $T$-dependent,
but it is assumed to be quite weak compared to $P(T)$.
Actually in the CO states in Fig.~\ref{Fig:PDexp}, there is no indication for 
noticeable changes of the spin order under fixed fields,
especially in the reentrant temperature region.\cite{Prellier} 
Thus the possible tiny change of
$\langle \cos (\theta /2) \rangle$ with $T$, even in the correct direction
(i.e., increasing with decreasing $T$),
is argued impossible for a melting of the CO, and thus not important. 
In what follows, $\langle \cos (\theta /2) \rangle$ will be simply
considered as a function of $H$, i.e., 
$\langle \cos (\theta /2) \rangle =F(H)$. 
Finally, a minor point is that, 
for the current model (\ref{H}),
no phonon-mediated intersite interaction has been derived. For other types
of e-p coupling, such an extra interaction may be obtained, leading to
a shift of the parameter $V$.\cite{Yuan,Yu}

For the effective Hamiltonian (\ref{Heff}), we make use of the Hartree 
mean-field (MF)
theory to study the CO. For a bipartite lattice with sublattices A and B,
it is assumed $\langle n_i \rangle = \left\{
\begin{array}{ll} 1/2+x, & i\in {\rm A}\\
1/2-x, & i\in {\rm B}\end{array}
\right.$, where $x$ characterizes the deviation from the uniform electron 
distribution. The final self-consistent equation for
the order parameter $x$ is obtained as follows (trival solution $x=0$
is ignored):\cite{Yuan}
\begin{eqnarray}
1 & = & {2V\over \pi^2}\int_{0}^{1}{\tanh [2\sqrt{(\tilde{t}z)^2+(Vx)^2}/T]
\over  \sqrt{(\tilde{t}z)^2+(Vx)^2}}K(\sqrt{1-z^2})\ {\rm d}z \ , \label{Eqx}
\end{eqnarray}
where $K$ is the complete elliptic integral of the first kind.

\section{Results and discussions}
The above equation can be easily evaluated numerically. For manganites,
the bare hopping $t$ is $\sim 0.2$-$0.5$ eV, and the intersite 
interaction $V/t$ is $\sim 0.1$-$0.2$ due to a large dielectric 
constant.\cite{Hotta}
The optical phonon frequency $\omega$ ranges $\sim 10$-$70$ meV. For example,
a soft mode at $\omega \simeq 7$ meV was reported by Zhao {\it et al.} for 
La$_{1-x}$Ca$_x$MnO$_3$ ($x=0.25,\ 0.4$) thin films.\cite{Guo} 
We point out that the high frequency modes play no role for the
melting of CO here because, as discussed above, 
they only contribute to a nearly flat
$P(T)$ in the temperature scale for the reentrance to occur.

\begin{figure}[ht]
\centerline{\epsfig{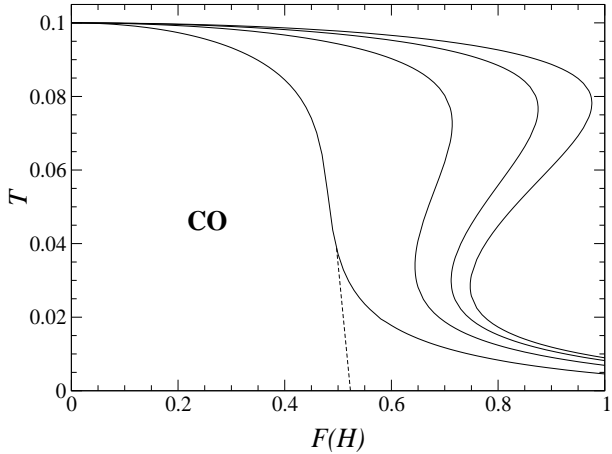}}
\smallskip
\caption{The phase diagram in the plane of $T$ and $F(H)$ 
for $V=0.1,\ \omega=0.05$. The solid curves from left to right correspond to
$g=0.035,\ 0.04,\ 0.042$, and $0.043$, respectively. The dashed line
shows a correction to the phase boundary of $g=0.035$ (see text).}
\label{Fig:PDw005}
\end{figure}

\begin{figure}[ht]
\centerline{\epsfig{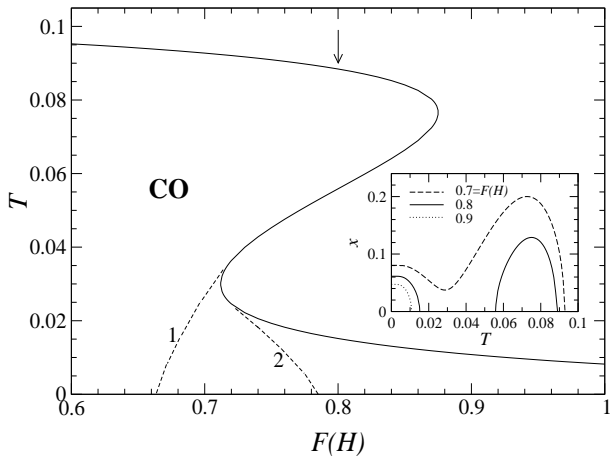}}
\smallskip
\caption{The enlarged phase diagram for $g=0.042$ in Fig.~\ref{Fig:PDw005}. 
The two dashed lines 1 and 2 show two possibilities for a correction to
the phase boundary (see text). In the inset
the order parameters $x$ vs. $T$ are plotted for $F(H)=0.7,\ 0.8$, and $0.9$.}
\label{Fig:PDg0042}
\end{figure}

The solution for $V=0.1,\ \omega=0.05$ is displayed in the phase 
diagram Fig.~\ref{Fig:PDw005} in the plane of $T$ and $F(H)$. The results
for different $g$ values are presented. The
reentrant CO is clearly seen within a finite region of $F(H)$ 
for $g=0.04,\ 0.042$, and $0.043$, while
not for $g=0.035$ for any $0\le F(H)\le 1$. For a careful analysis of the 
reentrant behavior,
the value $g=0.042$ is taken as an example and 
enlarged in Fig.~\ref{Fig:PDg0042}.
We fix $F(H)=0.8$ as indicated by the arrow and check how the CO
state evolves with decreasing $T$. 
At the highest energy scale, the entropic term dominates the free energy
and the homogeneous state is stable.
For temperatures of the order of $V$, where the coherent polaron
motion is nearly frozen (i.e., $\tilde{t}\rightarrow 0$), a charge order
transition takes place.\cite{SuppT}
On decreasing $T$ the band-type hopping $\tilde{t}$ becomes larger, 
and moreover, it increases fast enough to dominate $V$ so that
the CO state is destabilized. Thus the second transition, where the CO melts
into the homogeneous state (hs) takes place.
With continuing decrease of $T$, the third transition occurs to recover the
CO since the ground state must be charge ordered, which
will be discussed in detail later. To summarize, in the whole temperature
region, the system goes through a series of transitions: 
hs $\rightarrow$ CO $\rightarrow$ hs $\rightarrow$ CO with decreasing $T$.
The corresponding $x$ vs. $T$ is shown in the inset of Fig.~\ref{Fig:PDg0042}
by the solid line, where
the order parameter takes on alternating zero and non-zero values.
Interestingly, the double reentrance of the CO at fixed $F(H)$ 
or external field obtained here is a remarkable experimental 
observation shown in Fig.~\ref{Fig:PDexp}. In addition,
in the inset of Fig.~\ref{Fig:PDg0042}, the curves $x$ vs. $T$
for two examples without reentrance, i.e., $F(H)=0.7,\ 0.9$, are also shown.
An analysis of them is helpful. For $F(H)=0.7$, as shown by the dashed line,
$x$ decreases with decreasing
$T$ (due to increasing $\tilde{t}$) in the intermediate region 
but does not approach zero. This means that the increase of
$\tilde{t}$ is not sufficient to melt the CO.
The case for $F(H)=0.9$ is different. Now the absolute value of
$\tilde{t}$ becomes relatively large, even at high temperatures.
Although it is reduced at higher $T$, 
the simultaneously enhanced thermal fluctuation will prohibit CO. 
Thus the CO can not be stabilized at a 
relatively high temperature as the cases for $F(H)=0.7$ and $0.8$. 
It remains unstable with decreasing $T$, and appears 
only at very low temperatures to realize the charge ordered 
ground state, as seen by the dotted line.

As for the ground state, it is always charge ordered
for any $V/\tilde{t}>0$ because of the perfect nesting
property at half-filling for the model Hamiltonian (\ref{Heff}).
\cite{Gebhard,SuppHub}
This conclusion can be drawn from Eq.~(\ref{Eqx}),
which is proven to always give a solution of nonzero $x$ at $T=0$.\cite{Proof}
On the other hand, it must be noted that the perfect nesting property 
may be easily broken in real materials, for example, by inclusion of 
the next n.n. hopping which is usually present. 
In situations away from perfect nesting,
a natural conclusion for the ground state is that the CO only appears 
when $V$ exceeds a critical value, or equivalently it vanishes if $\tilde{t}$ 
becomes large.\cite{Kagan} Correspondingly, in the phase
diagram of Fig.~\ref{Fig:PDg0042}, a critical value of
$F(H)$ (or a critical field $H_c$) is expected at $T=0$ to 
separate the CO and homogeneous states.
With this in mind, the phase boundary may be reasonably corrected 
in two ways, 1 and 2,
as shown by the dashed lines. For line 1, only the possibility of
one reentrance is obtained. And for line 2, a double reentrance is 
still possible. These two possible phase diagrams are
qualitatively consistent with the experimental findings 
in Fig.~\ref{Fig:PDexp}.
At the same time, a similar correction is valid for the phase boundary 
of $g=0.035$ in Fig.~\ref{Fig:PDw005}, as shown by the dashed line.
This corrected phase boundary is the same as usually observed in 
half-doped manganites in the ($H,T$) plane, e.g, 
the bulk Pr$_{0.5}$Ca$_{0.5}$MnO$_3$.\cite{Tomioka,Tokunaga}
For concrete calculations including all above corrections,
a model beyond perfect nesting is needed, which is left for future studies.

As mentioned before, the reentrant CO on decreasing $T$ depends on 
the increasing strength of $\tilde{t}$ and its relative magnitude to $V$. 
Under which conditions it takes place is complicated, 
relying on the concrete parameters, 
i.e., $g,\ \omega, F(H)$ which control $\tilde{t}$, and the value of $V$. 
Some insight can be gained by comparing the phase diagrams with different 
values of $g,\ \omega,$ and $V$. 
A few consequences seem clear: for relatively large $F(H)$, increasing $g$ 
helps the first CO transition to occur at a high temperature, and consequently
favors the appearance of the reentrant behavior;
a similar effect can be obtained by increasing $V$; 
large $\omega$ is unfavorable to the reentrant CO.
Unfortunately, no quantitative criterion can be straightforwardly extracted.
\cite{Suppquant}

So far, we have not addressed the observation that the reentrant CO 
is exhibited in Pr(Nd)$_{0.5}$Ca$_{0.5}$MnO$_3$ thin films, 
while it is not obvious for the corresponding bulk
materials.\cite{Supp} The difference is understandable from the
above results. For thin films, due to strain effects, 
the lattice parameters are modified (e.g., the in-plane parameter is
elongated by the tensile strain). Correspondingly,
the model parameters will change in thin films compared to 
bulk materials.\cite{Comm} We speculate that the reentrant
behavior in thin films is due to the softened optical phonon modes.

Finally, we comment on the function $F(H)$, which was
used throughout the results. Its explicit form 
is not known although its qualitative behavior is transparent.
Above, spin order was fixed when the reduced Hamiltonian (\ref{H})
was introduced. Strictly speaking, the effect of $H$ should be studied 
interactively with the CO under the same Hamiltonian by inclusion of
the superexchange $J_{AF}$ of the core spins and the Zeeman energy.
The explicit form of $F(H)$ can be roughly estimated in the following 
phenomenological way. We consider a Hamiltonian of the form
$-t\langle \cos (\theta /2) \rangle \sum_{\langle ij \rangle}
(c_i^{\dagger}c_j + {\rm h.c.})
+J_{AF} \sum_{\langle ij \rangle} {\bf S}_i \cdot {\bf S}_j -
g\mu _B H\sum_i S_i^z\ ,
$
with ${\bf S}$: core spin and $g$: Land\'e factor. Moreover, a
uniform canted state from the competition between FM order due to the DE 
mechanism and AFM superexchange is assumed in order to 
simulate the essential effect of
the external field (i.e., favoring the FM order). 
In this way, a linear relation between $\langle \cos (\theta /2) \rangle$ 
and $H$ may be obtained.\cite{Zou}

\section{Conclusion}

Based on a double-exchange model with intersite Coulomb 
interaction and e-p coupling, charge ordering has been studied 
under magnetic field $H$ in the context of polaron formation. 
A melting of the CO on decreasing $T$ under fixed $H$ can be
naturally obtained due to the $T$-dependent band-type polaron hopping.
More generally, various phase diagrams in the plane of 
$T$ and $F(H)$ with (single or double) or without reentrant behavior, 
depending on the concrete model parameters, have been extensively discussed. 
The experimental findings on the CO under magnetic field 
in the half-doped Pr(Nd)$_{0.5}$Ca$_{0.5}$MnO$_3$ can
be explained for both thin films and bulk samples.
Future work will have to address a number of important issues  
which are still not well comprehended: the orbital degree
of freedom, a deviation from ($\pi,\pi$) for the wavevector 
of the CO in thin films,\cite{PreSim} and a possible magnetic 
phase separation in the CO state.\cite{Allodi}

\bigskip
We are very grateful to W. Prellier, R. Fr\'esard for valuable 
communications. This work was financially supported by the 
Deutsche Forschungsgemeinschaft through SFB 484 and the BMBF 13N6918/1. 
Q. Yuan also acknowledges P. Thalmeier for early collaborations and
the support by the National Natural Science Foundation of China.



\begin{references}
\bibitem{Salamon} For reviews, see M. B. Salamon and M. Jaime, Rev. Mod. Phys. 
 {\bf 73}, 583 (2001); A. P. Ramirez, J. Phys.: Condens. Matter {\bf 9}, 
 8171 (1997); J. M. D. Coey, M. Viret, and  S. von Molnar, 
 Adv. Phys. {\bf 48}, 167 (1999); W. Prellier, Ph. Lecoeur, and B. Mercey, 
 J. Phys.: Condens. Matter {\bf 13}, R915 (2001).
\bibitem{Maezono} R. Maezono, S. Ishihara, and N. Nagaosa, 
 Phys. Rev. B {\bf 58}, 11583 (1998).
\bibitem{Millis} A. J. Millis, Nature (London) {\bf 392}, 147 (1998);
 A. J. Millis, R. Mueller, and B. I. Shraiman, Phys. Rev. B {\bf 54}, 
 5405 (1996).
\bibitem{Horsch} P. Horsch, J. Jakli\v{c}, and F. Mack, 
 Phys. Rev. B {\bf 59}, 6217 (1999); R. Kilian and G. Khaliullin, 
 Phys. Rev. B {\bf 58}, R11841 (1998); {\bf 60}, 13458 (1999).
\bibitem{Rao} C. N. R. Rao, A. Arulraj, A. K. Cheetham, and B. Raveau,
 J. Phys.: Condens. Matter {\bf 12}, R83 (2000).
\bibitem{Imada} M. Imada, A. Fujimori, and Y. Tokura, Rev. Mod. Phys. 
 {\bf 70}, 1039 (1998).
\bibitem{PreGos} W. Prellier, A. M. Haghiri-Gosnet, B. Mercey, Ph. Lecoeur,
M. Hervieu, Ch. Simon, and B. Raveau, Appl. Phys. Lett. {\bf 77}, 1023 (2000).
\bibitem{PreSim} W. Prellier, Ch. Simon, A. M. Haghiri-Gosnet,
 B. Mercey, and B. Raveau, Phys. Rev. B {\bf 62}, R16337 (2000).
\bibitem{Ogimoto} Y. Ogimoto, M. Izumi, T. Manako, T. Kimura, Y. Tomioka,
 M. Kawasaki, and Y. Tokura, Appl. Phys. Lett. {\bf 78}, 3505 (2001). 
\bibitem{Biswas} A. Biswas, M. Rajeswari, R. C. Srivastava, T. Venkatesan, 
 R. L. Greene, Q. Lu, A. L. de Lozanne, and A. J. Millis, 
  Phys. Rev. B {\bf 63}, 184424 (2001).
\bibitem{Buzin} E. Rauwel Buzin, W. Prellier, Ch. Simon, S. Mercone,
 B. Mercey, and B. Raveau, Appl. Phys. Lett. {\bf 79}, 647 (2001).
\bibitem{Tokura} Y. Tokura and N. Nagaosa, Science {\bf 288}, 462 (2000).
\bibitem{Tomioka} Y. Tomioka, A. Asamitsu, H. Kuwahara, Y. Moritomo, and
  Y. Tokura, Phys. Rev. B {\bf 53}, R1689 (1996).
\bibitem{Tokunaga} M. Tokunaga, N. Miura, Y. Tomioka, and Y. Tokura, 
 Phys. Rev. B {\bf 57}, 5259 (1998). 
\bibitem{Respaud} M. Respaud, A. Llobet, C. Frontera, C. Ritter, 
 J. M. Broto, H. Rakoto, M. Goiran, and J. L. Garc\'ia-Mu\~noz,
 Phys. Rev. B {\bf 61}, 9014 (2000); F. Millange, S. de Brion, and 
 G. Chouteau, {\it ibid.} {\bf 62}, 5619 (2000).
\bibitem{Supp} From Refs. [\onlinecite{Tokunaga,Respaud}], a tiny reentrant 
 behavior is visible for bulk Nd$_{0.5}$Ca$_{0.5}$MnO$_3$, which, however,
 has no consequence to the theoretical results shown later. 
\bibitem{Jirak} Z. Jir\'{a}k, S. Krupi\v{c}ka, Z. \v{S}im\v{s}a, 
 M. Dlouh\'{a}, and S. Vratislav, J. Magn. Magn. Mater. 53, 153 (1985). 
\bibitem{Asami} Y. Tomioka, A. Asamitsu, H. Kuwahara, and Y. Tokura,
 J. Phys. Soc. Jpn. {\bf 66}, 302 (1997).
\bibitem{Khomskii} D. Khomskii, Physica B {\bf 280}, 325 (2000).
\bibitem{Kagan} M. Yu. Kagan, K. I. Kugel, and D. I. Khomskii, 
 JETP {\bf 93}, 415 (2001).
\bibitem{Kimura} T. Kimura, R. Kumai, Y. Tokura, J. Q. Li and Y. Matsui,
 Phys. Rev. B {\bf 58}, 11081 (1998); J. Q. Li, Y. Matsui, T. Kimura, 
 and Y. Tokura, {\it ibid.} {\bf 57}, R3205 (1998).    
\bibitem{Chatterji} T. Chatterji, G. J. McIntyre, W. Caliebe, 
 R. Suryanarayanan, G. Dhalenne, and A. Revcolevschi, Phys. Rev. B 
 {\bf 61}, 570 (2000).
\bibitem{Dho} J. Dho, W. S. Kim, H. S. Choi, E. O. Chi, and N. H. Hur,
  J. Phys.: Condens. Matter {\bf 13}, 3655 (2001).
\bibitem{Li} J. Q. Li, C. Dong, L. H. Liu, and Y. M. Ni, 
 Phys. Rev. B {\bf 64}, 174413 (2001).
\bibitem{Yuan} Q. Yuan and P. Thalmeier, Phys. Rev. Lett. {\bf 83}, 
  3502 (1999).
\bibitem{Koshibae} W. Koshibae, Y. Kawamura, S. Ishihara, S. Okamoto,
 J. Inoue, and S. Maekawa, J. Phys. Soc. Jpn. {\bf 66}, 957 (1997).
\bibitem{Mishra} S. K. Mishra, R. Pandit, and S. Satpathy, Phys. Rev. B
 {\bf 56}, 2316 (1997). 
\bibitem{Lee} J. D. Lee and B. I. Min, Phys. Rev. B {\bf 55}, R14713 (1997).
\bibitem{Yu} U. Yu, Yu. V. Skrypnyk, and B. I. Min, Phys. Rev. B 
 {\bf 61}, 8936 (2000).
\bibitem{Mutou} T. Mutou and H. Kontani, Phys. Rev. Lett. {\bf 83}, 
  3685 (1999).
\bibitem{Jackeli} G. Jackeli, N. B. Perkins, and N. M. Plakida, Phys. Rev. B 
 {\bf 62}, 372 (2000).
\bibitem{Cullen} J. R. Cullen and E. R. Callen, Phys. Rev. B
 {\bf 7}, 397 (1973).
\bibitem{Brink} J. van den Brink, G. Khaliullin, and D. Khomskii,
 Phys. Rev. Lett. {\bf 83}, 5118 (1999); {\bf 86}, 5843 (2001);
 S. Q. Shen, {\it ibid.} {\bf 86}, 5842 (2001).
\bibitem{Solovyev} I. V. Solovyev and K. Terakura, Phys. Rev. Lett. 
 {\bf 83}, 2825 (1999).
\bibitem{Yunoki} S. Yunoki, T. Hotta, and E. Dagotto, Phys. Rev. Lett. 
 {\bf 84}, 3714 (2000).
\bibitem{Hotta} T. Hotta, A. L. Malvezzi, and E. Dagotto, Phys. Rev. B 
 {\bf 62}, 9432 (2000).
\bibitem{LiUe} J. Q. Li, M. Uehara, C. Tsuruta, Y. Matsui, and Z. X. Zhao,
 Phys. Rev. Lett. {\bf 82}, 2386 (1999).
\bibitem{Zhao} G. M. Zhao, K. Ghosh, and R. L. Greene, J. Phys.: 
 Condens. Matter {\bf 10}, L737 (1998); G. M. Zhao, K. Ghosh, H. Keller,
 and R. L. Greene, Phys. Rev. B {\bf 59}, 81 (1999).
\bibitem{Alex} The polaronic carriers were also confirmed
 in extensive non-CO states of doped manganites, see 
 A. S. Alexandrov, G. M. Zhao, H. Keller, B. Lorenz, 
 Y. S. Wang, and C. W. Chu, Phys. Rev. B {\bf 64}, 140404(R) (2001) and 
 references therein.
\bibitem{Roder} H. R\"oder, J. Zang, and A. R. Bishop, Phys. Rev. Lett. 
 {\bf 76}, 1356 (1996).
\bibitem{Green} A. C. M. Green, Phys. Rev. B {\bf 63}, 205110 (2001).
\bibitem{Supp2D} Apparently, a more complicated model is needed to account
for the ``in-phase'' order along the $z$-axis, which is beyond the scope of
this paper. 
\bibitem{AlexBrat} A. S. Alexandrov and A. M. Bratkovsky, 
 J. Phys.: Condens. Matter {\bf 11}, 1989 (1999).
\bibitem{LeeMin} J. D. Lee and B. I. Min, Phys. Rev. B {\bf 55}, 12454 (1997).
\bibitem{Prellier} W. Prellier (private communication).
\bibitem{Guo} G. M. Zhao, V. Smolyaninova, W. Prellier, and H. Keller,
 Phys. Rev. Lett. {\bf 84}, 6086 (2000).
\bibitem{SuppT} The transition temperature is quantitatively
 overestimated by the MF theory. And it may be again corrected 
 by the incoherent processes, i.e., phonon states are altered 
 when an electron hops,\cite{Mahan} which are ignored in the derivation of
 Eq.~(\ref{Heff}) and become notable at high temperatures.
\bibitem{Mahan} G. D. Mahan, {\it Many Particle Physics} 
 (Plenum, New York, 1990), Sec. 6.2.
\bibitem{Gebhard} F. Gebhard, {\it The Mott Metal-Insulator Transition}
   (Springer-Verlag, Berlin, 1997), P218.
\bibitem{SuppHub} As a similar example, the 2D half-filled Hubbard model has
 AFM ground state for arbitrarily small on-site interaction.
\bibitem{Proof} The proof is obvious if three properties of the 
 right-hand side of Eq.~(\ref{Eqx}) are noticed (the $\tanh$ factor is 
 removed at $T=0$): (i). It is divergent at $x=0$. (ii). It decreases
 monotonically with increasing $x$. (iii). It is always lower than the
 function $1/(2x)$, which gives $1$ at $x=0.5$.
\bibitem{Suppquant} For real quantification, it'd better to first
 elaborate the formulas like Eq.~(\ref{Eqx}).
\bibitem{Comm} For example, the change of $t$ in thin films due to 
 purely geometric factors is discussed in Q. Yuan, to be published.
\bibitem{Zou} See also L. J. Zou, Q. Q. Zheng, and H. Q. Lin,
 Phys. Rev. B {\bf 56}, 13669 (1997).
\bibitem{Allodi} G. Allodi, R. De Renzi, F. Licci, and M. W. Pieper,
 Phys. Rev. Lett. {\bf 81}, 4736 (1998); F. Dupont, F. Millange, 
 S. de Brion, A. Janossy, and G. Chouteau, Phys. Rev. B {\bf 64}, 
 220403(R) (2001).
\end{references}
\end{document}